\documentclass[%
 reprint,
 superscriptaddress,
 preprintnumbers,
 amsmath,amssymb,
 aip,
]{revtex4-1}

\usepackage[utf8]{inputenc}
\usepackage{graphicx}
\usepackage{grffile}
\usepackage[usenames,dvipsnames]{xcolor}
\usepackage{amsmath}
\usepackage[normalem]{ulem}
\usepackage[resetlabels, labeled]{multibib}
\usepackage{appendix}
\newcites{S}{References Supplementary Materials}
\definecolor{orange}{rgb}{1,0.5,0}
\definecolor{goodgreen}{rgb}{0.1,0.5,0}
\definecolor{goodred}{rgb}{0.7,0,0}
\usepackage{lineno}
\setpagewiselinenumbers
\modulolinenumbers[5]

\usepackage[colorlinks,urlcolor=goodgreen,citecolor=blue,linkcolor=goodred]{hyperref}
\usepackage{float}
\usepackage{babel}
\graphicspath{ {images/} }

\makeatother

\let\oldepsilon\epsilon \let\epsilon\varepsilon \let\varepsilon\oldepsilon
\let\oldphi\phi \let\phi\varphi \let\varphi\oldphi

\begin{document}

\title{Bipolar thermoelectricity in S/I/NS and S/I/SN superconducting tunnel junctions}

\newcommand{\orcid}[1]{\href{https://orcid.org/#1}{\includegraphics[width=8pt]{orcid.png}}}

\author{A. Hijano}
\altaffiliation{Authors to whom correspondence should be addressed: \href{mailto:alberto.hijano@ehu.eus}{alberto.hijano@ehu.eus}, \href{mailto:fs.bergeret@csic.es}{fs.bergeret@csic.es}, \href{mailto:francesco.giazotto@sns.it}{francesco.giazotto@sns.it}, \href{mailto:alessandro.braggio@nano.cnr.it}{alessandro.braggio@nano.cnr.it}}
\affiliation{Centro de F\'isica de Materiales (CFM-MPC) Centro Mixto CSIC-UPV/EHU, E-20018 Donostia-San Sebasti\'an,  Spain}
\affiliation{Department of Condensed Matter Physics, University of the Basque Country UPV/EHU, 48080 Bilbao, Spain}
\author{F. S. Bergeret}
\altaffiliation{Authors to whom correspondence should be addressed: \href{mailto:alberto.hijano@ehu.eus}{alberto.hijano@ehu.eus}, \href{mailto:fs.bergeret@csic.es}{fs.bergeret@csic.es}, \href{mailto:francesco.giazotto@sns.it}{francesco.giazotto@sns.it}, \href{mailto:alessandro.braggio@nano.cnr.it}{alessandro.braggio@nano.cnr.it}}
\affiliation{Centro de F\'isica de Materiales (CFM-MPC) Centro Mixto CSIC-UPV/EHU, E-20018 Donostia-San Sebasti\'an,  Spain}
\affiliation{Donostia International Physics Center (DIPC), 20018 Donostia--San Sebasti\'an, Spain}
\author{F. Giazotto}
\altaffiliation{Authors to whom correspondence should be addressed: \href{mailto:alberto.hijano@ehu.eus}{alberto.hijano@ehu.eus}, \href{mailto:fs.bergeret@csic.es}{fs.bergeret@csic.es}, \href{mailto:francesco.giazotto@sns.it}{francesco.giazotto@sns.it}, \href{mailto:alessandro.braggio@nano.cnr.it}{alessandro.braggio@nano.cnr.it}}
\affiliation{NEST Istituto Nanoscienze-CNR and Scuola Normale Superiore, I-56127 Pisa, Italy}
\author{A. Braggio}
\altaffiliation{Authors to whom correspondence should be addressed: \href{mailto:alberto.hijano@ehu.eus}{alberto.hijano@ehu.eus}, \href{mailto:fs.bergeret@csic.es}{fs.bergeret@csic.es}, \href{mailto:francesco.giazotto@sns.it}{francesco.giazotto@sns.it}, \href{mailto:alessandro.braggio@nano.cnr.it}{alessandro.braggio@nano.cnr.it}}
\affiliation{NEST Istituto Nanoscienze-CNR and Scuola Normale Superiore, I-56127 Pisa, Italy}



%


\begin{abstract}
Recent studies have shown the potential for bipolar thermoelectricity in superconducting tunnel junctions with asymmetric energy gaps. The  thermoelectric performance of these systems is significantly impacted by the inverse proximity effects present in the normal-superconducting bilayer, which is utilized to adjust the gap asymmetry in the junction. Here, we identify the most effective bilayer configurations, and we find that directly tunnel-coupling the normal metal side of  the cold bilayer with the hot superconductor is more advantageous compared to the scheme used in experiments. By utilizing quasiclassical equations, we examined the nonlinear thermoelectric junction performance as a function of the normal metal film thickness and the quality of the normal-superconducting interface within the bilayer, thereby determining the optimal design to observe and maximize this nonequilibrium effect. Our results offer a roadmap to achieve improved thermoelectric performance in superconducting tunnel junctions, with promising implications for a number of applications.

\end{abstract}

\maketitle

Recent theoretical and experimental works have reported a sizeable thermoelectric effect in conventional S/I/S' tunnel junctions, where S and S' are superconductors with different energy gaps~\cite{Marchegiani:2020a,Marchegiani:2020b, Marchegiani:2020c,Germanese:2021,Germanese:2022,Germanese:2023,Bernazzani:2022} separated by a thin insulating tunnel barrier (I). 
 Several different thermoelectric elements have been so far proposed with superconductors, for instance,  superconducting-ferromagnetic systems\cite{Bergeret:2018,heikkila2019thermal,machon2013nonlocal,ozaeta2014predicted,kolenda2016observation,Heikkila:2018,Chakraborty:2018,Heidrich:2019,strambini2022superconducting,Geng:2020,Geng:2022}, in which the electron-hole symmetry is broken by the combination of spin-splitting and polarization of the barrier or by other phase-coherent\cite{Titov:2008,Jacquod:2010,Kalenkov:2017,Dolgirev:2018,Kalenkov:2021-Hybrids,Kalenkov:2021-Andreev,Blasib:2020} or non-local effects\cite{Claughton:1996,Eom:1998,Virtanen:2007,Mazza:2014,Hussein:2019,Blasi:2020,Tan:2021}.
Yet, the thermoelectric effect in SIS' junctions, when the Josephson coupling is suppressed, relies on spontaneous particle-hole (PH) symmetry breaking induced by the strong non-equilibrium condition, i.e., the large temperature difference imposed across the system. Intriguingly, since the PH symmetry of the lead density of states (DoS) determines a full reciprocal $I(-V)=-I(V)$ characteristics, the thermoelectricity, signalled by $V I(V)<0$, is necessarily \emph{bipolar}. This means that two opposite thermoelectric voltages/currents are equivalently generated with the \emph{same} thermal gradient thereby realizing a unique functionality for a thermoelectric device.~\cite{Marchegiani:2020a} The absolute negative resistance in superconducting junctions with different gaps was predicted in~\cite{Aronov:1975} and experimentally observed in~\cite{Gershenzon:1986,Gershenzon:1988}. Recent experiments~\cite{Germanese:2022,Germanese:2023} exploited a normal-superconducting (NS) bilayer for fine tuning 
the asymmetry between the junction gaps
via the inverse proximity effect.  
However, inverse proximity will also affect the sharpness of the DoS, thus negatively impacting the thermoelectric generation, as recent measurements seem to indicate\cite{Germanese:2023}.\\
Here, we  theoretically investigate how different configurations of the NS bilayer tunnel junctions  affect the nonlinear bipolar thermoelectricity. Specifically, we compare the thermoelectric performance between S/I/SN and S/I/NS cases. 
We find that the latter configuration, counterintuitively, promises a much improved thermoelectric performance than the configuration adopted in the experiments so far. Finally, we investigate the junction thermoelectric response as a function of the N film thickness, the SN interface resistance, and the hot lead temperature. \\ 
%
%
\begin{figure}[!t]
  \includegraphics[width=0.99\columnwidth]{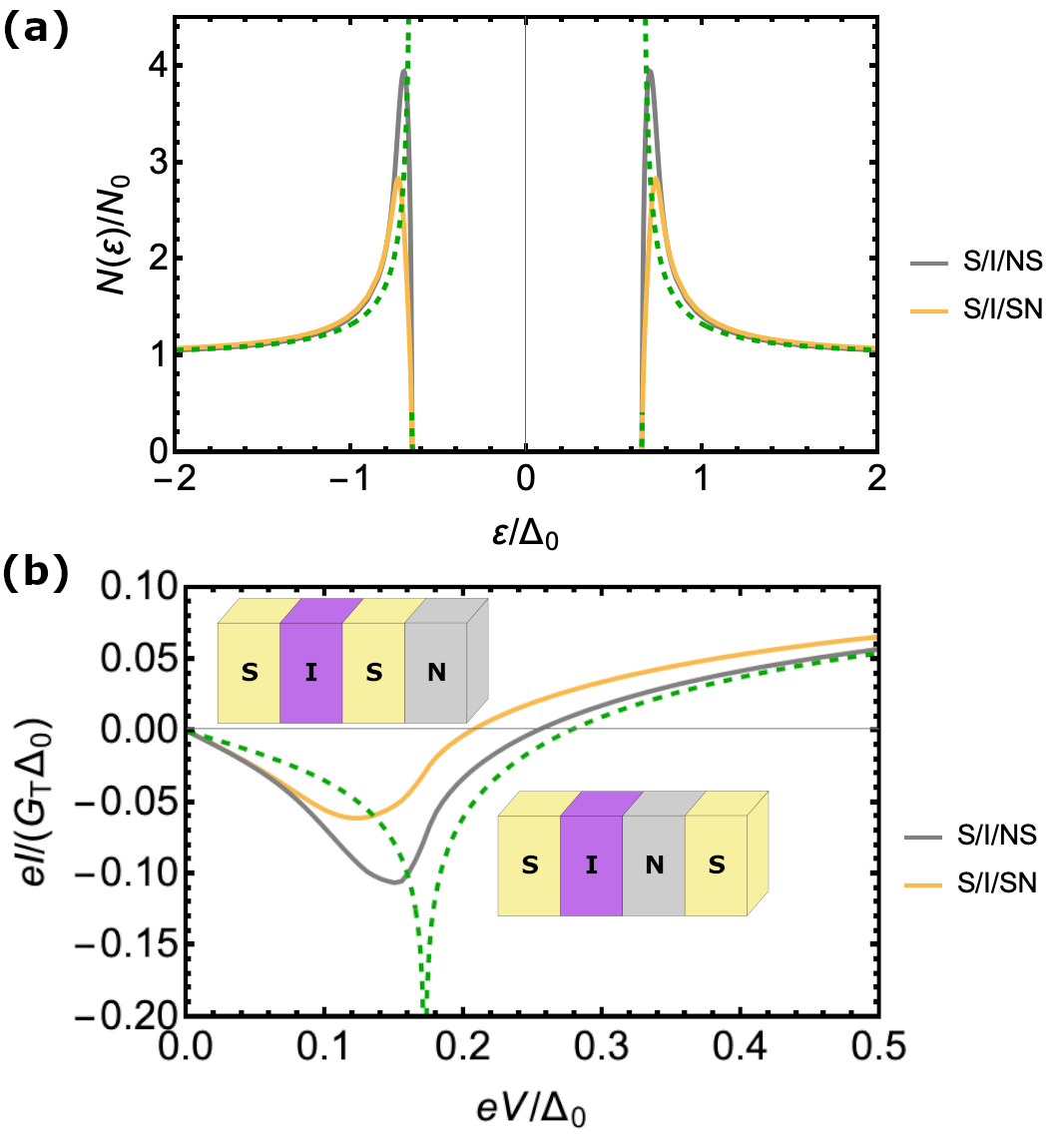}
\caption{(a) Density of states of the NS bilayer at the outer interface for the S (yellow) and N (gray) layers. The green dashed line corresponds to a bulk superconductor with the same effective gap. (b) Current vs voltage characteristic of a S/I/SN (yellow) and S/I/NS (gray) junctions. The thickness of the S and N layers are $d_S=\xi_0$ and $d_N=0.1\xi_0$. The temperatures of the left and right electrodes are $T_1=0.7T_{c0}$ and $T_2=0$.}\label{Fig:1a}
\end{figure}
%
%
To compute the transport properties for generic configurations, in the tunneling limit, we will use the quasiclassical Green’s function (GF) formalism\cite{Belzig:1999}. Because we are dealing with superconductivity, the quasiclassical Green's function $\check{g}$ is described by a $2 \times 2$ matrix in Nambu space $\check{g}=g\tau_3+f\tau_1$, where $\tau_i$ are the Pauli matrices in Nambu space, and $g$ and $f$ are the normal and anomalous parts of the GF. In dirty systems, the mean free path $\ell$ is smaller than the superconducting coherence length ($\ell\ll\xi_0$) with $\xi_0=\sqrt{\hbar D/\Delta_0}$, where $D$ is the diffusion coefficient. In such systems, the quasiclassical equations reduce to a diffusive-like equation known as the Usadel equation~\cite{Usadel}
\begin{equation}\label{usadel_equation}
    D \partial_x(\check{g}\partial_x\check{g})+[i(\epsilon+i\Gamma)\tau_3-\Delta\tau_1,\check{g}]=0\; ,
\end{equation}
where $\epsilon$ indicates the energy and $\Gamma$ is the Dynes parameter~\cite{Dynes:1978} describing inelastic scattering.  The superconducting gap $\Delta$ is determined self-consistently~\cite{Kopnin:2001} by the gap equation 
\begin{equation}
\label{self-consistency}
\Delta\ln\left(\frac{T}{T_{c0}}\right)=2\pi T\sum_{n=0}\left(f(\omega_n)-\frac{\Delta}{\omega_n}\right)
\end{equation}
where $\omega_n=2\pi T(n+1/2)$, $n \in \mathbb{Z}$, are the Matsubara frequencies, 
$T_{c0}$ is the zero-field critical temperature and $f(\omega_n)$ is the Matsubara anomalous GF, obtained by analytic continuation of the GF to the complex plane $\epsilon \rightarrow i\omega_n$. In writing Eq.~\eqref{usadel_equation} we have assumed that all junctions are translational invariant in the $(y,z)$-plane and hence the GF only depends on $x$. The Usadel equation~\eqref{usadel_equation} is supplemented by boundary conditions describing the interfaces between different materials. The spectral current vanishes at the boundaries with vacuum or an insulator, in the GF language this boundary condition translates into $\check{g}\partial_x\check{g}=0$ at the outer interfaces of the NS bilayer. The NS interface is described by the Kupriyanov-Lukichev boundary condition~\cite{Kupriyanov-Lukichev}
\begin{equation}
   \check{g}_N\partial_x\check{g}_N=\check{g}_S\partial_x\check{g}_S=\frac{1}{2\xi_0\rho}\left.[\check{g}_N,\check{g}_S]\right|_{x=0}\; .
\end{equation}
The interface quality is described by the dimensionless parameter $\rho=\sigma_N R_{_\square}/\xi_0$, where $\sigma_N$ is the normal-state conductivity and $R_{_\square}$ is the NS interface resistance per unit area.\\
We assume in the following that the Josephson coupling between the two sides of the junction is suppressed, for instance,  by applying a suitable in-plane magnetic field to induce Fraunhofer interference or via a small out-of-plane magnetic field in a superconducting quantum interference device (SQUID)~\cite{Germanese:2022,Germanese:2023}. In such limit, the $IV$ characteristics is dominated by the quasiparticle component $I_{\mathrm{qp}}$ for the tunneling current~\cite{Barone_Paterno,Josephson:1962,Harris:1974,Gulevich:2017}
\begin{equation}
    I_{\mathrm{qp}}=\frac{1}{eR}\int^{\infty}_{-\infty}\!\!\mathrm{d}\epsilon \mathcal{N}_{1}(\epsilon+eV)\mathcal{N}_{2}(\epsilon)\left(f_2(\epsilon)-f_1(\epsilon+eV)\right),
\end{equation}
where the normalized DoS of the $i$th electrode is $\mathcal{N}_i(\epsilon)=\mathrm{Re}\{g_i(\epsilon)\}/N_0$, with $N_0$ being the normal-state DoS. The lead electron distributions are Fermi-like, $f_i(\epsilon)=(\mathrm{e}^{\epsilon /k_\mathrm{B} T_i}+1)^{-1}$, since we assume that the electrons on the two sides of the barrier are respectively in thermal equilibrium at temperature $T_i$ with $i=1,2$.
For nonequilibrium $T_1\neq T_2$, there is the possibility to develop the bipolar thermoelectric effect and, in the following, we will compare different junction configurations with the target to maximize the 
junction thermoelectric performance.
%
%
\begin{figure*}[!t]
  \includegraphics[width=0.99\textwidth]{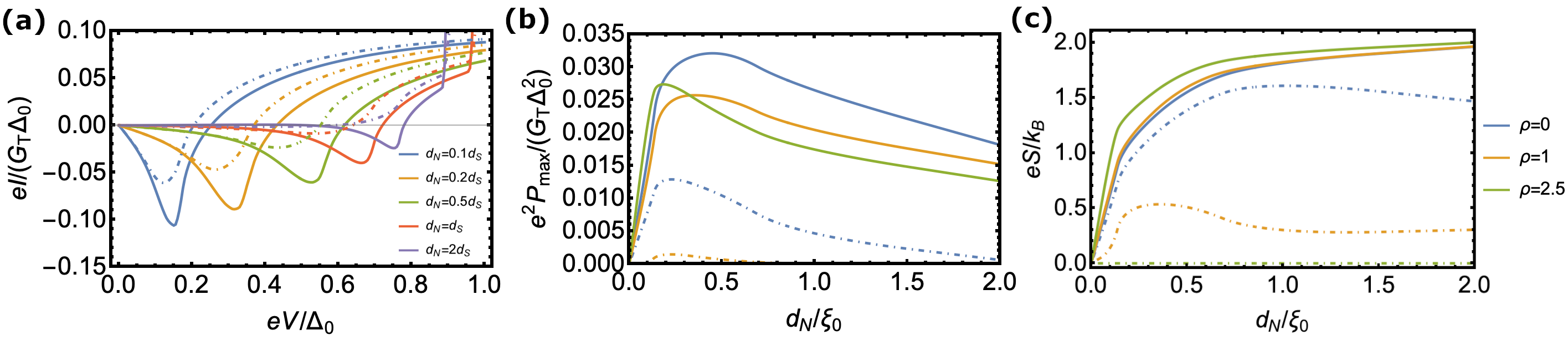}
\caption{(a) Thermoelectric current vs voltage characteristic curves of a S/I/NS (solid) and S/I/SN (dot-dashed) junctions for different thicknesses of the N layer. (b) Maximum thermoelectric power and (c) Seebeck coefficient as a function of the thickness of the N layer for different interface SN resistances $\rho$ for S/I/NS (solid) and S/I/SN (dot-dashed) junctions. The temperatures of the left and right electrodes are the same of Fig.~\ref{Fig:1a}  and the thickness of the cold S layer is kept fixed to $d_S=\xi_0$.}\label{Fig:2}
\end{figure*}
%
%

The crucial ingredients for the nonlinear thermoelectricity of a superconducting tunnel junction are leads with different energy gaps and a sufficiently strong thermal gradient.~\cite{Marchegiani:2020a,Marchegiani:2020b,Marchegiani:2020c} In particular, for SIS', the temperature of the superconductor with the largest gap ($T_1$) must be sufficiently high $T_1\gtrsim T_2/r$ with respect to the temperature of the superconductor with the smallest gap ($T_2$)~\cite{Marchegiani:2020a,Germanese:2022}, where $r=\Delta_{0,2}/\Delta_{0,1}<1$ is the asymmetry parameter of the two zero-temperature gaps $\Delta_{0,i}$ with $i=1,2$.\\ 
Asymmetric S/I/S' junctions can be achieved by using different superconducting materials for each electrode. Alternatively, the same superconducting material can be utilized by taking advantage of the inverse proximity effect, where a normal layer (N) is attached to one of the S leads, allowing for the controlled suppression of the gap~\cite{Germanese:2022}.
In the following, we compare the thermoelectric effect in S/I/NS and S/I/SN junctions, since these are very relevant configurations for experimental applications [see inset in Fig.~\ref{Fig:1a}(b)].
Different configurations of junctions can exhibit different thermoelectric performance due to the proximity effect, which affects the DoS differently depending on which side of the interfaces is contacted. To address this issue, we first investigate how the DoS of an NS bilayer depends on the contacted side of the interface by solving Eq.~\eqref{usadel_equation} for an SN bilayer with thicknesses $d_S=\xi_0$ and $d_N=0.1\xi_0$, assuming a perfect NS interface where the N layer is proximitized by an S layer and vice versa. We consider a thick superconducting film where the superconductivity still survives but with a reduced gap $\Delta< \Delta_0$ in the regime where the DoS is no longer spatially homogeneous and exhibits smearing in energy. Figure~\ref{Fig:1a}(a) shows the DoS of the metal (in gray) and the superconducting (in yellow) sides of the NS bilayer, as well as the DoS of a bulk Bardeen-Cooper-Schrieffer (BCS) superconductor with an equivalent gap (dashed green line) and the same Dynes parameter $\Gamma=10^{-4}\Delta_0$.
The DoS at the N side is much more peaked than that at the S layer, consistent with previous results obtained for thin SN bilayers with an intermediate-valued interface resistance~\cite{Fominov:2001} and in FI/N/S junctions~\cite{Silaev:2020}. This sharp peaks at energies slightly smaller than the superconducting gap are attributed to quasiparticle excitations of de Gennes Saint-James bound states~\cite{de_Gennes_Saint-James:1963}. When an electron in the N region with energy lower than $\Delta$ is reflected at the NS interface it forms a coupled electron-hole bound state with energy lower than $\Delta$. In thin N layers, the bound states remain at energies close to $\Delta$, resembling the BCS peaks of a bulk superconductor~\cite{Wolf:1982}.
As thermoelectricity strongly depends on the energy dependence of the DoS, this suggests that the N side of the bilayer may perform better than the S side in making the tunneling junction, which is counterintuitive.\\
To clarify this point, we compare in Figure~\ref{Fig:1a}(b) the nonlinear thermoelectric current-voltage characteristic curves of two tunneling junctions, S/I/NS (in gray) and S/I/SN (in yellow), with the same $d_N$ and $d_S$. The left and right electrodes are subjected to a strong temperature gradient, with the left superconductor at $T_1=0.7T_{c0}$ and the bilayer, which has the smaller gap, at $T_2=0$. We assume zero temperature for simplicity, but experiments show that similar temperature gradients can be realized. Given the gap ratios $r\approx 0.8$, the nonlinear bipolar thermoelectricity is expected to be generated. The dashed green line shows the thermoelectric characteristic of a generic S/I/S' junction where S' has a gap $\Delta_2$ equivalent to the NS bilayer but with a standard BCS DoS. At subgap voltages $e|V|\lesssim\Delta_1+\Delta_2$, the current $I_{\mathrm{qp}}$ flows against the bias ($V I_{\mathrm{qp}}(V)<0$), and the junction is thermoelectric in this voltage range. The thermocurrent $|I_{\mathrm{qp}}|$ reaches a maximum around the matching peak $eV_p=\Delta_1-\Delta_2$, and for $e|V|\gtrsim\Delta_1+\Delta_2$, the junction becomes dissipative. \\
The thermoelectric performance of the system is strongly influenced by the DoS at the tunneling interface of the electrodes. Therefore, the orientation of the cold NS bilayer plays a crucial role in maximizing the thermoelectric current. The results in panel (a) suggest that the matching peak is more spread out in the S/I/SN configuration compared to the S/I/NS configuration. Furthermore, the matching peak is deeper and broader in the latter. From these observations, we conclude that the S/I/NS configuration is more suitable for thermoelectricity, as evidenced by the two most straightforward figures of merit, namely, the Seebeck thermovoltage, $V_S$, for which $I_{\mathrm{qp}}(V_S)=0$, and the maximum thermocurrent, $I_\mathrm{max}$, which is given by $\operatorname{max}_{0<V<V_S}(|I_{\mathrm{qp}}(V)|)$. This conclusion is somewhat counterintuitive since only the S/I/SN configuration has been used in previous experiments.~\cite{Germanese:2022,Germanese:2023} However one may speculate that this is not generally true but connected to the specific value of the parameters adopted or that other non-universal parameters can potentially affect this result, but we will see that it is not the case.\\
%
%
\begin{figure}[!t]
  \includegraphics[width=0.99\columnwidth]{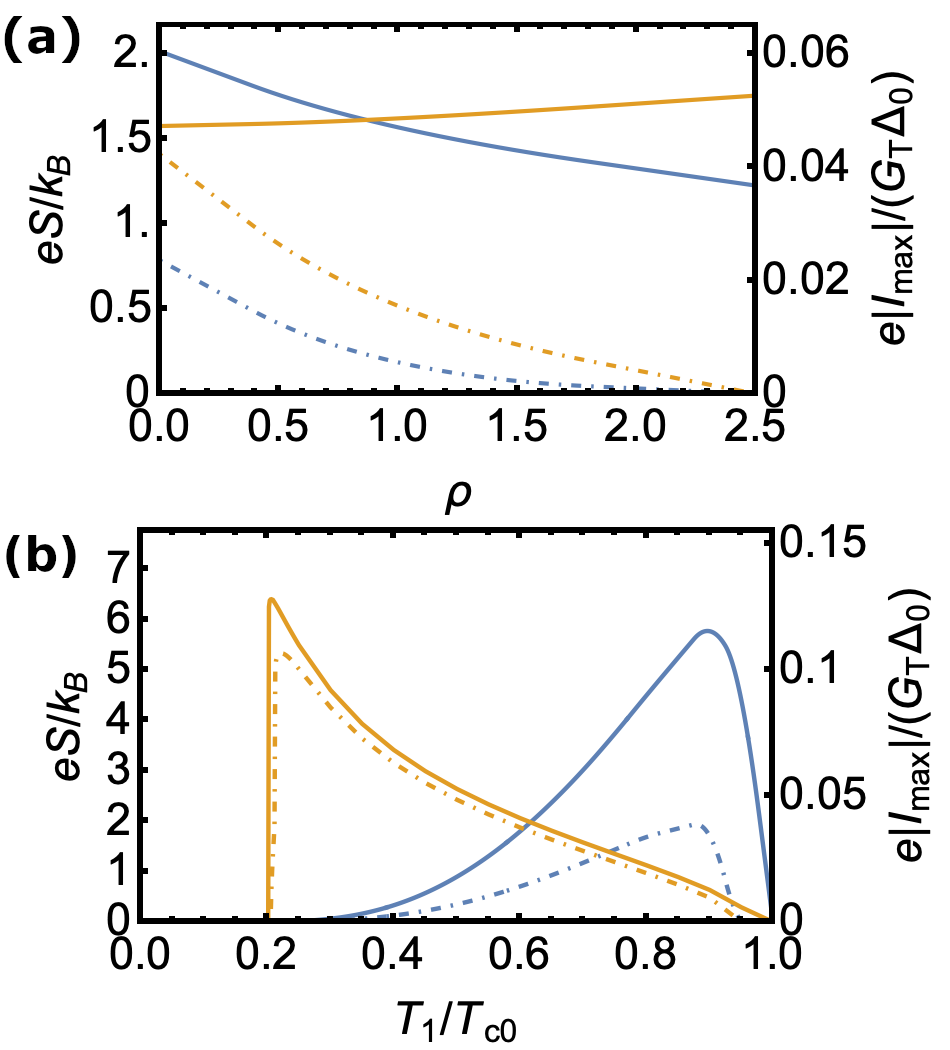}
\caption{Nonlinear Seebeck coefficient $S$ (yellow, left scale) and maximal thermocurrent $I_{\mathrm{max}}$ (blue, right scale)  for the S/I/NS (solid) and S/I/SN (dot-dashed) junctions, as a function of the NS interface quality parameter $\rho$ for $T_1=0.7T_{c0}$ (a), and  as a function of the hot temperature $T_1$ for $\rho=0$  (b). The parameters used are $d_S=\xi_0$, $d_N=0.5\xi_0$, $T_2=0$.}\label{Fig:3}
\end{figure}
%
%
It is crucial to investigate the impact of the thickness of the N layer ($d_N$) on the thermoelectric properties of both types of junctions. This is illustrated in Fig.~\ref{Fig:2}(a). For very thin N layers, the effective gap of the NS bilayer $\Delta_2$ is somewhat similar to that of the S layer so the matching peak $V_p$ takes a small value. At the same time $\Delta_2$ decays monotonously with increasing $d_N$, so the matching peak is progressively shifted to higher voltages in thicker samples. The same happens to the Seebeck thermovoltage $V_S$ typically $V_S\gtrsim V_p$ at least until when the N thickness is so large that there is no-more thermoelectricity overall. Indeed, for the S/I/SN case (dot-dashed), thicker $d_N$ induces a bigger gap renormalization $\Delta_2$ and, at some point, the junction even ceases completely to be thermoelectric being the gap asymmetry too big.~\cite{Marchegiani:2020a} Instead, for the S/I/NS junction, there is also potentially another competitive mechanism: the leakage of Cooper pairs from the S to the N layer which occurs over a characteristic length scale $\xi_N$ away from the SN interface. In a diffusive normal metal, without magnetic impurities, this length is of the order of the thermal length $\xi_N=\sqrt{\hbar D/k_B T}$. So the N DoS at distances larger than $\xi_N$ will rapidly evolve into the normal state DoS resembling more the S/I/N junction where no thermoelectricity is expected neither in the linear nor in the nonlinear temperature regime.~\cite{Marchegiani:2020a}
Those mechanisms determine that the thermoelectric current $|I_{\mathrm{qp}}(V_p)|$ at the matching peak is typically reduced with thicker N layers almost independently of the configuration.  
Notably, we observe that  S/I/SN junctions (dot-dashed line) consistently under-perform relative to  S/I/NS junctions for any value of $d_N$, further confirming the previously discussed general statement.\\
The discussed mechanisms will also determine the peculiar non-monotonic behavior of the maximum thermoelectric power $P_\mathrm{max}=\operatorname{max}_{0<V<V_S}(-V I_{\mathrm{qp}}(V))$. 
Figure~\ref{Fig:2}(b) shows  the maximum electric power output of the junction, $P_\mathrm{max}\approx -V_p I_{\mathrm{qp}}(V_p)$, as a function of $d_N/\xi_0$. We note that there is no thermoelectricity, i.e., $P_\mathrm{max}\to 0$ for $d_N\to 0$ and $d_N\to \infty$. For S/I/NS junctions (solid), in the first limit, there is no proximity gap renormalization so the junction loses the gap asymmetry, whereas in the second limit, the junction becomes an S/I/N, and there is no thermoelectricity. 
We see that, in general, the maximal power is strongly reduced for the S/I/SN junctions (dot-dashed). In the limit $d_N\to \infty$, the bilayer gap $\Delta_2$ is so strongly suppressed that thermoelectricity disappears altogether.
\\
Furthermore, it is interesting to note that the non-linear Seebeck coefficient $S=V_S/(T_1-T_2)$ exhibits different behavior with respect to $d_N$ for the two junction types. As shown in Fig.~\ref{Fig:2}(c), for S/I/NS junctions, the non-linear Seebeck coefficient increases by increasing $d_N$. However, for S/I/SN junctions, the non-linear Seebeck coefficient has a maximum value at intermediate thicknesses and then progressively decreases for $d_N\to \infty$. 
This difference between the two configurations in the limit $d_N\gg \xi_0 $ is a notable consequence of the non-linearity of the thermoelectric effect reported here. Indeed, in a linear regime, under the thermal gradient $\Delta T$, the maximal thermoelectric power is $P_\mathrm{max}=G S^2\Delta T^2 /4$ which depends on $S$ and $G$, i.e., the junction Seebeck coefficient and electrical conductance, respectively.~\cite{Benenti:2017}
It is interesting now to compare in Fig.~\ref{Fig:2}(b) and (c) the curves for $P_\mathrm{max}$ and the nonlinear Seebeck coefficient $S$, respectively, corresponding to different NS interface resistance $\rho$.
We start discussing the S/I/NS configuration (solid line). For ideal interfaces ($\rho=0$), layers where $d_N \approx 0.5\xi_0$ optimize the generated thermoelectric power. For finite interface resistance $\rho>0$ the leakage of Cooper pairs into the N layer is hindered, which reduces the depth of the matching peak and the generated thermoelectric power. This shifts the optimal thickness to a slightly smaller value. 
However, we observe that for very small $d_N$, a non-perfect SN interface could be even slightly better. Nevertheless, in general, a worse quality of the SN interface is detrimental to the nonlinear thermoelectric performance. 
A similar behavior is observed for the S/I/SN configuration (dot-dashed line), even if the optimal thickness is typically smaller. However, for this configuration, we find that the quality of the interface is even more crucial than the S/I/NS case. This provides an additional experimental reason to prefer the normal-metal side of the NS bilayer.\\ 
Figure~\ref{Fig:3}(a)  displays how  $I_\mathrm{max}$ and the nonlinear Seebeck $S$ coefficient depends on the NS interface quality $\rho$. The maximal current (blue lines) decreases by increasing $\rho$ in both junctions due to the suppression of the proximity effect. The Seebeck coefficient (yellow lines) shows opposite behavior in each junction configuration. In the S/I/SN junction, an increase of $\rho$ results in a higher effective gap $\Delta_2$ on the superconducting region. In this case, the matching peak $eV_p=\Delta_1-\Delta_2$ is shifted to lower voltages, and the Seebeck voltage is reduced similarly. By contrast, in the S/I/NS junction an increase of $\rho$ hinders the proximity effect on the normal metal, reducing the effective gap on the N layer, so that the Seebeck voltage is shifted to higher values.\\
Figure~\ref{Fig:3}(b) shows the two main figures of merit discussed previously as a function of the hot side temperature $T_1$. We see that the behavior of these two quantities generally agrees with the overall behavior reported for the S/I/S' junctions.~\cite{Marchegiani:2020a,Marchegiani:2020b}
On the one hand, it is observed that the best Seebeck coefficient $S$ is obtained at the lowest hot lead temperatures, although they still need to be above a threshold value so that thermoelectricity may arise. On the other hand, the maximal thermocurrent is obtained at $T_1\approx 0.9 T_c$, which is slightly higher than the optimal hot temperature reported for S/I/S' junctions.  This suggests that bilayer-based systems may have a slightly higher thermopower than junctions made with different superconductors with identical gaps.  Moreover, the difference between the two configurations is more significant for the thermopower than for the Seebeck thermovoltage and coefficient.  This implies that the two configurations are quite equivalent for applications where the thermovoltage is relevant, such as current-controlled thermoelectric memories\cite{Germanese:2022,Patent1} or single-photon sensors.\cite{Paolucci:2023,Patent2,Hijano:2022-microwave} However, S/I/NS junctions are better suited for thermoelectric engines\cite{Marchegiani:2020b,Marchegiani:2020b} and energy harvesting, where the thermopower is the relevant figure of merit.

We have presented  a comprehensive investigation of the influence of the proximity effect of the SN bilayer on the nonlinear bipolar thermoelectric effect by using the quasiclassical  Green's function method for diffusive systems. In particular, we have compared the commonly used S/I/SN configuration with the S/I/NS configuration, and have discovered that, contrary to expectations, the latter generally exhibits improved thermoelectric performance, regardless of the thickness of the normal metal film, the quality of the NS interface, and the temperature of the hot lead. Our findings yield a set of general design principles and rules for creating superconducting thermoelectric tunnel junctions using a bilayer NS film technology, which can aid in maximizing the thermoelectric performance for various applications.
We expect that similar conclusions may be relevant also for the cooling performance of S/I/S' electron refrigerators \cite{Giazotto:2006,quaranta2011cooling} and transistors \cite{giazotto2005josephson,tirelli2008manipulation} realized with NS bilayers.\\

A.H. acknowledges funding from the University of the Basque Country (Project PIF20/05). F.S.B. acknowledges  financial support from Spanish AEI through projects PID2020-114252GB-I00 (SPIRIT) and TED2021-130292B-C42, the Basque Government through grant IT-1591-22, and the A. v. Humboldt Foundation. F.G. and A.B. acknowledge EU’s Horizon 2020 Research and Innovation Framework Programme under Grant No. 964398 (SUPERGATE), No. 101057977 (SPECTRUM) and the PNRR MUR project PE0000023-NQSTI. A.B. acknowledges the MIUR-PRIN2022 Project NEThEQS (Grant No. 2022B9P8LN) and the Royal Society through the International Exchanges between the UK and Italy (Grants No. IEC R2 192166.).\\
\section*{References}
\vspace*{-\baselineskip}
\nocite{*}
%

\end{document}